\documentclass[11pt,a4paper]{article}
\usepackage{amsfonts,amsmath,amstext,amssymb}
\usepackage{dsfont}
\usepackage{theorem}    %para cambiar el estilo de los teoremas
\usepackage{a4wide}

\newtheorem{lema}{Lemma}[section]

\newtheorem{teor}[lema]{\bf Theorem}

\newtheorem{rema}[lema]{\bf Remark}

\theorembodyfont{\rmfamily}

\newenvironment{demo}{\noindent \textit{Proof.} }{\par \hfill{$\Box$}}

\title{Complete constant mean curvature spacelike hypersurfaces in the Einstein-de Sitter spacetime}

\author{Rafael M. Rubio\footnote{The author is partially supported by the
Spanish MEC-FEDER Grant MTM2010-18099.}
\\[0.5mm] Departamento de Matem\'aticas, Campus de Rabanales, \\[0.5mm]
Universidad de C\'ordoba, 14071 C\'ordoba, Spain,\\[0.5mm]
E-mail\textup{:\texttt{\;rmrubio@uco.es}}}

\date{}

\begin{document}

\maketitle

\begin{abstract}Uniqueness and non-existence results on complete constant mean curvature spacelike hypersurfaces lying between two spacelike slice in the Einstein-de Sitter spacetime are given. They are obtained from a Liouvielle-type theorem applied to a distinguished smooth function on a constant mean curvature hypersurface.

\end{abstract}

\thispagestyle{empty}

\vspace{1mm}

\noindent {\it PACS:} 02.40Ky, 04.20Cv, 04.20.Jb\\
\noindent {\it Keywords:} Spacelike hypersurface, Constant mean curvature, Einstein-de Sitter spacetime.

\hyphenation{ma-ni-fold}

\section{Introduction}

The Einstein-de Sitter spacetime is a classical exact solution to the Einstein field equation without cosmological constant. It is a open Fiedmann-Robertson-Walker model, which incorporated homogeneity and isotropy (the cosmological principle) and permitted expansion. Moreover, the Einstein-de Sitter universe has showed a reasonable fit to recent observations (Type Ia supernovae, hubble space telescope) \cite{V}.

Our main aim in this paper is to give new uniqueness results for complete contant mean curvature spacelike hypersurfaces in the classical Einstein-de Sitter spacetime.

The importance in General Relativity of maximal and constant mean curvature spacelike hypersurfaces in spacetimes is well-known; a summary of several reasons justifying it can be found in \cite{M-T}. In particular hypersurfaces of constant mean curvature (non-zero) are particularly suitable for studying the propagation of gravity radiation \cite{S}. Classical papers dealing with uniqueness are \cite{Ch}, \cite{BF} and \cite{M-T}, although a previous relevant result in this direction was the proof of the Bernstein-Calabi conjecture \cite{Calabi} for n-dimensional Minkowski spacetime given by Cheng and Yau , \cite{Cheng-Yau}. In \cite{BF}, Brill and Flaherty replaced Minkowski spacetime by a spatial closed universe, and proved uniqueness in the large by assuming ${\mathrm{Ric}}(z,z)>0$ for all timelike vector $z$. In \cite{M-T}, this energy condition was relaxed by Marden and Tipler to include, for instance, non-flat vacuum spacetimes. More recently Bartnik, proved in \cite{Bar} very general existence theorems and consequently, he claimed that it would be useful to find new satisfactory uniqueness results. Still more recently, in \cite{A-R-S1} Alias, Romero and Sanchez prove new uniqueness results in the class that the authors call spatially closed generalized Robertson-Walker spacetimes (which includes the spatially closed Robertson-Walker spacetimes), under Temporal Convergence Condition, in \cite{A-M} Al\'{\i}as and Montiel, improve these results and using a generalized maximum principle due to Omori \cite{OM} an Yau \cite{YA}, they give also (see Therem 10) a general uniqueness result for the case of complete constant mean curvature spacelike hypersurfaces whose Ricci curvature is bounded from below, working this results in the paticular case of the Einstein-De Sitter spacetime. Finally, in \cite{RRS}, Romero , Rubio and Salamanca, has given uniqueness results in the maximal case, for spatially parabolic generalized Robertson-Walker spacetimes, which are open model, whose fiber must be a parabolic Riemannian manifold.

The purpose of this work  is to establish in a different approach of \cite{A-M}, uniqueness and non existence  results of complete constant mean curvature spacelike hypersurfaces (without any assumption on its Ricci curvature) in the Einstein-De-Sitter spacetime. In fact, we define on a spacelike hypersurface, the hyperbolic angle function $\theta$, between its normal unitary vector field and the oposite to the timelike coordinate vector field. Making use of a Liouvielle-type theorem (see Theorem 3.1, section 3) we prove that $\theta$ vanishes identically. On the other hand, the function $\theta$ admits a reasonable physical interpretation (see section 2). Clearly the spacelike slices of the Einstein-de Sitter spacetime have constant mean curvature, thus, we pose the following natural question: When is a complete spacelike hypersurface of constant mean curvature a spacelike slice?

The contents of this paper are organized as follow. In the section 2, we introduce the notation and several formulas to be used later. In section 3, we obtain a central inequality and recall the aforementioned
Liouvielle-type theorem, which will be used into the principal results. Finally, in section 4, we show the uniqueness and non existence announced results.

\section{Preliminaries}
The Einstein-de Sitter spacetime is given by the product manifold $\overline{M}=(0,\infty)\times\mathbb{R}^3$, endowed with the Lorentzian metric
\begin{equation}\label{metric}
\overline{g}(\,,\,)= -dt^ 2+ t^{2/3}(dx_1^2+dx_2^ 2+dx_3^ 2).
\end{equation}
\noindent  This is, a warped product in the sense of \cite[pag. 204]{O-N} with base the open interval $(0,\infty)$ endowed with the negative metric, fiber the Euclidean space $(\mathbb{R}^3,\langle\,,\,\rangle)$ and warping function $f(t)=t^{1/3}$. We denote $\pi_I$ and $\pi_F$  the projections onto $(0,\infty)$ and $\mathbb{R}^3$

On $\overline{M}$ there is a distinguished
vector field $K: =({\pi}_I)^{1/3}\,\partial_t,$ which is timelike and,
from the relationship between the Levi-Civita connections of $\overline{M}$
and those of the base and the fiber \cite[Cor. 7.35]{O-N}, it
satisfies
\begin{equation}\label{conexion}
\overline{\nabla}_XK =\frac{1}{3}{\pi}_I^{-2/3} \,X,
\end{equation}
for any $X\in \mathfrak{X}(M)$, where $\overline{\nabla}$ is the
Levi-Civita connection of the metric (\ref{metric}). Thus, $K$ is
conformal with $\mathcal{L}_{K} \, \overline{g}
=\frac{2}{3}{\pi}_I^{-2/3}\,\overline{g}$ and its metrically
equivalent 1-form is closed.

Given an $3$-dimensional manifold $S$, an immersion $x: S
\rightarrow \overline{M}$ is said to be \textsl{spacelike} if the Lorentzian
metric given by (\ref{metric}) induces, via $x$, a Riemannian
metric $g$ on $S$. In this case, $S$ is called a spacelike
hypersurface. The Lorentzian manifold $(\overline{M},\overline{g})$ is
clearly time orientable. This allows us to take, for each
spacelike hypersurface $S$ in $\overline{M}$,  $N \in \mathfrak{X}^\bot(S)$
as the only globally defined unitary timelike vector field normal
to $S$ in the same time-orientation of the vector field $-
\partial_t$, $\partial_t:=\partial/\partial_t$, i.e., such that
$\overline{g}(N,-\partial_t)<0$. From the wrong-way Cauchy-Schwarz
inequality (see \cite[Prop. 5.30]{O-N}, for instance), we have
$\overline{g}( N,
\partial_t) \geq 1$, and the equality holds at a point $p\in S$ if
and only if $N_p = -\partial_t$. In fact, $\overline{g}( N,
\partial_t)=\cosh \theta$, where $\theta$ is the hyperbolic angle,
at any point, between the unit timelike vectors $N$ and
$-\partial_t$. We will refer to $\theta $ as the
\textsl{hyperbolic angle} between $S$ and $-\partial_t$.  
The function $\theta$ has a physical interpretation.
In fact, consider the unitary normal vector field $N$ on $S$ and the
unit timelike vector field $\mathcal{T}_{p}:=-\partial_t$ (the sign
minus depends on the chosen time orientation). Along $S$ there exist
two families of \emph{instantaneous observers} $\mathcal{T}_{p}$,
$p\in S$, and the normal observers $N_p$. The quantities $$\cosh
\theta (p) \quad \text{and} \quad
v(p):=\left(\frac{1}{\cosh\theta(p)}\right)\, N_p^F,$$ where
$N_p^F$ is the projection of $N_p$ onto $\mathbb{R}^3$ and $\theta$ the
hyperbolic angle of $S$, are respectively the \emph{energy} and the
\emph{velocity} that $\mathcal{T}_p$ measures for $N_p$, and we have
$\vert v\vert= \tanh\theta$ on $S$, \cite[pp. 45,67]{Sa-Wu}.

 If we denote by
$\partial_{t}^{\top}:=\partial_{t}\,+\overline{g}(N,\partial_t)N$
the tangential component of $\partial_{t}$ along $x$, and
by $\tau:=\pi_I \circ x$, then it is not difficult to obtain the
following formula for its gradient,
\begin{equation}\label{gradiente}
\nabla \tau = -\partial_t^\top,
\end{equation}
and therefore
\begin{equation}\label{modulo_gradiente}
g(\nabla \tau, \nabla \tau) = \sinh^2 \theta \, .
\end{equation}

Let us represent by $\nabla$ the Levi-Civita connection of the
metric $g$. The Gauss and Weingarten formulas of $S$ are
respectively written
\begin{equation}\label{Gauss}
\overline{\nabla}_X Y = \nabla_X Y - g(AX,Y )N \, ,
\end{equation}
\begin{equation}\label{Weingarten}
A X = - \overline{\nabla}_X N \, , \qquad \qquad \quad
\end{equation} for all $X, Y \in \mathfrak{X}(S)$,
where $A$ is the shape operator associated to $N$. Recall that the
\textsl{mean curvature function} relative to $N$ is $H:= -(1/3)
\mathrm{trace}(A)$. The mean curvature is zero if and only if the
spacelike hypersurface is, locally, a critical point of the
3-volume functional for compactly supported normal
variations. A spacelike hypersurface with $H=0$ is called a
\textsl{maximal} hypersurface.

\vspace{1mm}

In the Einstein-de Sitter spacetime $\overline{M}$, the level hypersurfaces of the function
$\pi_I :\overline{M} \longrightarrow I$ constitute a distinguished family of spacelike
hypersurfaces: the so-called \textsl{spacelike slices}. Along this paper, we will 
represent by $t=t_0$ the spacelike slice $\{t_0\}\times \mathbb{R}^3$. For a given spacelike hypersurface  
$x : S \longrightarrow \overline{M}$, we have that $x(S)$ is contained in $t=t_0$ if and only if
$\pi_I \circ x=t_0$ on $S$. We will say that $S$ is a spacelike slice if $x(S)$ equals to
$t=t_0$, for some $t_0\in I$ , and that $S$ is contained between two
slices if there exist $t_1,t_2 \in I$, $t_1 < t_2$, such that
$x(S) \subset \left[t_1, t_2 \right] \times \mathbb{R}^3 \,$.

\vspace{1mm}

The shape operator and the mean
curvature of the spacelike slice $t=t_0$ are respectively
$A=f'(t_0)/f(t_0)\,I$, where $I$ denotes the identity
transformation, and the constant $H= - f'(t_0)/f(t_0)$. Thus,
a spacelike slice $t=t_0$ is maximal if and only if $f'(t_0)=0$ (and
hence, totally geodesic). Obviously, there is no maximal slices in the Einstein-De Sitter spacetime.  
\vspace{1mm}

Now, given a spacelike hypersurface $S$ in $\overline{M}$, from
(\ref{conexion}) and the Gauss and Weingarten formulas
(\ref{Gauss}), (\ref{Weingarten}) we get
\begin{equation}\label{derivada_covariante}
\nabla_Y K^\top + (\tau)^{1/3} \, \overline{g}(N,\partial_t) \, AY =
\frac{1}{3}(\tau)^{-2/3}\, Y \, ,
\end{equation}for any $Y\in\mathfrak{X}(S)$, where $K^\top = K + \overline{g}(K,N) N \,$
is the tangential component of $K$ along $x$. Taking into account that
$K^\top=\tau^{1/3} \, \partial_t^T$ and (\ref{gradiente}), by
contraction of (\ref{derivada_covariante}) we arrive to

\begin{equation}\label{lapla}
\Delta \tau = - \frac{1}{3\tau} \left\{ 3 + |\nabla
\tau|^2 \right\} - 3 H \, \overline{g}(N, \partial_t) \, ,
\end{equation}
where $\Delta$ denotes the Laplacian on $S$.

\vspace{1mm}

As a consequence

\begin{equation}\label{lapla2}
\Delta \tau^ {1/3} = - \frac{1}{3}{\tau^{-5/3}} -\frac{1}{3}{\tau^{-5/3}} |\nabla \tau|^2-\tau^ {-2/3}H\cosh \theta .
\end{equation}

Let us consider  the function $\overline{g}(N, K)=\tau^{2/3}\cosh\theta$ on $S$. It is immediate to see that 
\begin{equation}\label{gradient}
\nabla\overline{g}(N, K)=-AK^\top,
\end{equation}

\noindent where $K^\top:=K+\overline{g}(N, K)N$ is the tangential component of $K$ on $S$.

A direct computation from (\ref{gradient}) gives

\begin{equation}\label{gradient2}
\nabla\cosh\theta=-A\partial _t^\top + \frac{2}{3\tau}\overline{g}(N,\partial_t)\partial_t^\top.
\end{equation}

\noindent Using (\ref{gradient2})
we have

\begin{equation}\label{lapl}
\Delta\overline{g}(N, K)=\Delta(\tau^{1/3}\cosh\theta)=-\frac{1}{3}\tau^{-5/3}\cosh\theta-\frac{1}{3}\tau^{-5/3}\cosh\theta\sinh^2\theta-\tau^{-2/3}H\cosh^2\theta
\end{equation}

$$+\tau^{1/3}\Delta\cosh
\theta + 2g(A\partial_t^\top,\partial_t ^\top)-\frac{2}{9}\tau^{-5/3}\cosh\theta\sinh^2\theta.$$

 On the other hand, if the hypersurface $S$ has constant mean curvature, using  the Gauss and Weingarten formulas, as well as the Codazzi equation, we have

\begin{equation}\label{lapla}
\Delta\overline{g}(N, K)=\overline{{\rm Ric}}(K^\top, N)+\tau^{-2/3}H+\tau^{1/3}\cosh\theta\, {\rm trace}(A^2),
\end{equation}

\noindent where $\overline{{\rm Ric}}$ stands for the Ricci tensor of the spacetime.

\section{Setup}

If we put $N=N_{_F}-\overline{g}(N,\partial_t^\top)\partial_t$, where $N_{_F}$ denotes the lift of the projection of the vector field $N$ on the fiber $\mathbb{R}^3$, it is easy to obtain from \label{metric}
$$\sinh\theta=\tau^{2/3}\langle N_{_F},N_{_F}\rangle.$$

On the other hand, taking into account \cite[Cor. 7.43]{O-N}, we obtain

\begin{equation}
\overline{\rm Ric}(K^\top,N)=\frac{2}{3}\tau^{-5/3}\cosh\theta\sinh^2\theta
\end{equation}

\noindent Now, from (\ref{lapl}) and (\ref{lapla}) we get

\begin{equation}\label{laplacosh}
\Delta\cosh\theta=\frac{1}{\tau}H(1+\cosh^ 2\theta)+\frac{1}{\tau^ 2}\sinh^ 2\theta\cosh\theta+\frac{1}{9}{\tau^ {-5/3}}\cosh\theta\,(3+2\sinh^ 2\theta)
\end{equation}
$$+\cosh^2\theta\,{\rm trace}(A^2)-\frac{2}{3\tau}\cosh\theta\, g(A\partial_t^\top,\partial_t^\top).\,\,\,\,\,\,\,\,\,\,\,\,\,\,\,\,\,\,\,\,\,\,\,\,\,$$

On the other hand, the square algebraic trace-norm of the Hessian tensor of $\tau$ is just

$$\vert {\rm Hess}(\tau)\vert^2={\rm trace} (H_\tau\circ H_\tau),$$

\noindent where the operator defined by $g({\rm Hess}_\tau(X),Y):={\rm Hess}(\tau)(X,Y)$, for all $X,Y$ vector fields on $S$.

If we take tangential component in (\ref{conexion}) and we use (\ref{gradiente}), we get that

\begin{equation}\label{hess}
\vert {\rm Hess}(\tau)\vert=\frac{1}{9\tau^2}\{2+\cosh^4\theta\}+\cosh^2\theta\,{\rm trace}(A^2)+\frac{2}{\tau}H\cosh\theta-\frac{2}{3\tau}\cosh\theta\,g(A\partial_t^\top,\partial_t^\top).
\end{equation}

Making use of (\ref{hess}) in (\ref{laplacosh}), we obtain

\begin{equation}\label{acot}
\cosh\theta\Delta\cosh\theta\geq\frac{1}{\tau}H\cosh\theta\sinh^ 2\theta+\frac{11}{9\tau^ 2}\cosh^ 2\sinh^ 2\theta+\frac{1}{3\tau^ 2}\cosh^ 2\theta-\frac{1}{9\tau^ 2}\{2+\cosh^ 4\theta\}.
\end{equation}

Now, we recall the following Liouvielle-type theorem \cite{Cheng-Yau}, \cite{N}, which will be a fundamental tool in our work.

\begin{teor}\label{lema}
Let $\Sigma$ be a complete Riemannian manifold whose Ricci curvature is bounded from below and let $u:\Sigma\longrightarrow\mathbb{R}$ be a non-negative smooth function on $\Sigma$. If there exists a constant $c>0$ such that $\Delta u\geq cu^2$, then $u$ vanishes identically on $\Sigma$.
\end{teor}

\section{The principal results}

\begin{lema}\label{ricci}  Let $S$ be a constant mean curvature  spacelike hypersurface in the Einstein-de Sitter spacetime $(\overline{M},\overline{g})$, then its Ricci curvature is bounded from below. 
\end{lema}

\begin{demo} Indeed, if we denote by Ric the Ricci curvature of $S$, it is no difficult to see
$${\rm Ric}(X,X)\geq \Sigma_{i=1}^3 g(\overline{{\rm R}}(X,E_i)X,E_i)-\frac{9H^2}{4}\vert X\vert^2,$$

\noindent for all $X\in\mathfrak{X}(S)$, where $\{E_1,E_2,E_3\}$ denotes a local orthonormal frame on an open $\Theta\subset S$ and $\overline{{\rm R}}$ the curvature tensor of the Einstein-de Sitter spacetime.

If we put $\partial_t^\top=\partial_t +\overline{g}(N, \partial_t)N$ and analogously we split the vector field $X$ as the addition of its projections on the base and the fiber, using \cite[Prop. 7.42]{O-N}, we obtain 

$$\Sigma_{i=1}^3 g(\overline{{\rm R}}(X,E_i)X,E_i)\geq \frac{2}{9\tau^2}\vert X\vert^2+\frac{2}{3\tau^2}g(X,\partial_t^\top)^2+\frac{1}{3\tau^2}\vert\partial_t^\top\vert^2\vert X\vert^2,$$
\noindent thus ${\rm Ric}(X,X)\geq -\frac{9H^2}{4}\vert X\vert^2$.
\end{demo}

\begin{teor} There is no complete maximal hypersurfaces in the Einstein-de Sitter spacetime, which lie between two spacelike slices.
\end{teor}

\begin{demo} Consider the function $\sinh^2\theta$ on $S$ and observe that
$$\Delta\sinh^2\theta =\Delta\cosh^2\theta=2\cosh\theta\Delta\cosh\theta + 2\vert\nabla\cosh\theta\vert^2.$$

From (\ref{acot}), we have 
\begin{equation}\label{des}
\cosh\theta\Delta\cosh\theta\geq \sinh^2\theta(\frac{3}{2}H+\frac{1}{3\tau}\cosh\theta)^2+3\sinh^2\theta(\frac{1}{9\tau^2}-\frac{3}{4}H^2)+\frac{1}{9\tau^2}\sinh^ 4\theta.
\end{equation}
\noindent Since $S$ lies between two slices, there exists a constant $c>0$ such that

$$\Delta\sinh^2\theta\geq c\sinh^4\theta.$$

Now, the Theorem \ref{lema} is called and we can to conclude that $\sinh\theta=0$, i.e., $S$ is a spacelike slice, which is contradictory.
\end{demo}

\vspace{2mm}

As a  consequence of a trivial modification in the proof of \cite[Corollary 5.3]{C-R-R}, we know that on every complete  spacelike hypersurface with constant mean curvature $H$ in the Einstein-De Sitter spacetime, which is contained between two slices, holds the inequality 

$$H^2\leq \frac{1}{9\tau^2}.$$

\noindent Thus, reasoning as in the previous result, we  obtain (compare with \cite[Theorem 10]{A-M}).

\begin{teor} The only complete constant mean curvature spacelike hypersurfaces in the Einstein-de Sitter spacetime, which lie between two spacelike slices are spacelike  slices.
\end{teor}

\begin{rema} Observe that the Lemma \ref{ricci} and the inequality (\ref{des}) can be easily extended to another  Robertson-Walker spacetimes with flat fiber, which obey the Timelike Convergence Condition, i.e., its warping function $f$ satisfies $f''\leq 0$.
\end{rema}

\end{document}